\documentclass[11pt]{article}
\usepackage{amssymb,amsmath,amsthm,latexsym,natbib,amsbsy,mathtools}
\usepackage{graphicx}
\usepackage{float}
\usepackage{caption}
\usepackage{subcaption}
\usepackage{epsfig}
\usepackage{epstopdf}  
\usepackage{color}
\usepackage[font= {small,it} ]{caption}
\usepackage{url}
\usepackage{doi}
\usepackage[titletoc]{appendix}
\usepackage{xcolor}  
\usepackage[boxed]{algorithm2e} 

\usepackage[margin= 1in]{geometry}

\newcommand{\bs}{\boldsymbol}
\renewcommand{\a}{\alpha}
\renewcommand{\b}{\beta}

\newcommand{\Y}{\boldsymbol Y}

\author{ {\sc Stella Watson Self$^1$, Christopher McMahan$^{1,*}$, D. Andrew Brown$^1$,}\\ \sc{ Robert Lund$^1$, Jenna Gettings$^2$, and Michael Yabsley$^{2,3}$} \\[.2in]
$^1$Department of Mathematical Sciences \\ Clemson University \\ Clemson, SC 29634-0975 \\
$^2$Southeastern Cooperative Wildlife Disease Study \\Department of Population Health \\ The University of Georgia \\ Athens, GA 30602 \\
$^3$Warnell School of Forestry and Natural Resources \\ The University of Georgia \\ Athens, GA 30602 \\[.2in]
$^*$ \emph{mcmaha2@clemson.edu}}

\title{ {\bf A Large Scale Spatio-temporal Binomial Regression Model for Estimating Seroprevalence Trends}}
\date{}

\begin{document}

\maketitle
\thispagestyle{empty}

\begin{abstract}
This paper develops a large-scale Bayesian spatio-temporal binomial regression model for the purpose of investigating regional trends in antibody prevalence to \textit{Borrelia burgdorferi}, the causative agent of Lyme disease. The proposed model uses Gaussian predictive processes to estimate the spatially varying trends and a conditional autoregressive model to account for spatio-temporal dependence. Careful consideration is made to develop a novel framework that is scalable to large spatio-temporal data. The proposed model is used to analyze approximately 16 million \textit{Borrelia burgdorferi} test results collected on dogs located throughout the conterminous United States over a sixty month period. This analysis identifies several regions of increasing canine risk. Specifically, this analysis reveals evidence that Lyme disease is getting worse in some endemic regions and that it could potentially be spreading to other non-endemic areas. Further, given the zoonotic nature of this vector-borne disease, this analysis could potentially reveal areas of increasing human risk.

\end{abstract}

\newpage

\section{Introduction}

Lyme disease is a vector-borne disease that impacts both humans and several other mammalian species, with domestic dogs being particularly sensitive to infection \citep{Little2010}. Disease occurs as a result of infection by \textit{Borrelia burgdorferi}, a spirochetal bacteria that is transmitted by ticks. Incidence of disease in humans is considered to be emerging, with a growing number of high incidence counties \citep{Adams2017}. Humans and dogs are infected by the same vectors \citep{Little2010}, and so, unsurprisingly, the risks of exposure for both are closely related. In fact, dogs are often considered to be sentinels for the regional risk of Lyme disease in humans \citep{Mead2011}.

Dogs are tested regularly for exposure to \textit{B.~burgdorferi} as part of their annual wellness examinations. Commonly, veterinarians use a serologic test that detects antibodies against the C6 peptide that is present in the blood of infected animals. The presence of C6 is indicative of an intermediate or late-term infection, and is often detectable 3 to 6 weeks after exposure \citep{Wagner2012}. Among dogs that are infected, only approximately 5\% develop any clinical signs of Lyme disease \citep{Levy1992}. This practice of routine testing provides a unique opportunity to measure the seroprevalence of \textit{B.~burgdorferi} within a relatively healthy canine population visiting veterinary clinics.

Monitoring seroprevalence is useful for many reasons, despite the low incidence of disease. Directly, it provides an estimate for the risk of exposure within a region, allowing veterinarians to make accurate preventative care and testing recommendations. Indirectly, the seroprevalence of \textit{B.~burgdorferi} can identify the approximate range of the \textit{Ixodes} spp. tick vectors. Especially because \textit{Ixodes} spp. are capable of transmitting several other pathogens, including \textit{Anaplasma}, \textit{Ehrlichia muris eauclairensis} and \textit{Babesia microti} \citep{Nelder2016}, several of which are zoonotic.  The shared vector allows dogs to serve as sentinels for human risk. Therefore, modeling trends in canine seroprevalence should inform changing risk of exposure to \textit{B. burgdorferi} in humans.

The goal of this paper is to identify US regions that are experiencing an increase in canine seroprevalence of \textit{B.~burgdorferi}, and by proxy identify regions where the risk of human exposure could also be increasing. The data analyzed here contain 16,571,562 serologic test results for \textit{B.~burgdorferi} conducted on domestic dogs in the conterminous United States (US) from January 2012 - December 2016, aggregated by county and month. Figure 1 displays the raw prevalence estimates after aggregating over all sixty months in the study; i.e., the proportion of positive tests. There are 3,109 distinct US counties and county-equivalent regions in the conterminous US, not all of which report test data every month.  Data were reported from 69,876 county-month pairs.  As our goal is to determine {\em where} seroprevalence is increasing, our model must have a temporal trend component that is spatially varying. To facilitate more reliable inference, the strong positive spatio-temporal dependence of the tests is also taken into account. The size of this data set and its large spatio-temporal support motivates some of our methodological choices.

Gaussian processes (GPs) are popular geostatistical modeling tools due to their flexibility and ability to quantify uncertainty in nonparametric regressions \citep{OHagan78, Neal98}. Overviews of GP modeling can be found in \citet{Cressie93}, \citet{Rasmussen06}, \citet{Cressie11}, and \citet{Gelfand16}. \citet{BanerjeeEtAl04} discuss Bayesian aspects of GPs. Objective prior specification for GP models is studied in \citet{Berger01}.  GPs have become standard tools in a wide variety of applications, including oceanography \citep{Jona-Lasinio13}, water quality analysis \citep{Zhang08}, image classification \citep{Morales-Alvarez17}, neuroimaging \citep{Lazar08}, and computer experiments \citep{SantnerEtAl03}. GPs are also used to model disease prevalence, including dengue fever \citep{Johnson17}, Malaria \citep{Andreade-Pacheco15}, and influenza \citep{Senanayake16}. \citet{Gelfand03} used GPs to allow linear model coefficients to vary smoothly over space, an approach used here to localize regional trends in \textit{B.~burgdorferi} seroprevalence.

Gaussian process modifications and algorithms for analyzing big spatial data sets have received significant attention in the recent literature, including fixed rank kriging \citep{CressJohn08} and LatticeKrig \citep{NychkaEtAl15} approaches.  Both methods employ basis function expansions of spatial random effects to reduce the dimension of the associated covariance matrices. \citet{Katzfuss17} took a similar approach by applying basis functions to a succession of refined resolutions. Spatial partitioning \citep[e.g.,][]{SangEtAl11, HeatonEtAl17} can be used to split regions into smaller, more manageable sub-regions with computation being accelerated via a conditional independence assumption. Covariance tapering \citep{FurrerEtAl06} uses a covariance function with compact support to induce sparsity. Nearest neighbor processes \citep{DattaEtAl16} achieve computational efficiency by conditioning on a subset of nearby observations. A similar idea was used by \citet{GramacyApley15} to find the largest number of neighbors computationally feasible for prediction, optimally chosen by minimizing prediction variance. \cite{LGPReview17} provide an overview and comparison of these procedures and others. The approach used here involves Gaussian predictive processes (GPPs) \citep{Banerjee08}, which are discussed further in Section \ref{Sec:Model}.

The most common approach for modeling spatially dependent areal data involves Gaussian Markov random fields \citep[GMRFs;][]{Rue05}, with Gaussian conditional autoregressive (CAR) models \citep{BanerjeeEtAl04} being particularly popular. As special cases of Markov random fields \citep{Besag74}, GMRFs are collections of jointly distributed Gaussian random variables satisfying a Markov dependence structure quantified through a precision matrix. GMRFs are extended to flexible degrees of smoothness in \citet{Brezger07} and \citet{Yue08}. \citet{BrownEtAl17} adjust the CAR precision matrix to build a unified model for independent and dependent cases and study neighborhood structures other than those based on physical adjacency.  GMRF and GP connections are explored in \citet{RueTjem02}, \citet{SongEtAl08}, and \citet{Lindgren11}.  CAR models are now standard in disease mapping problems \citep[e.g.,][]{Waller97}.

To achieve our goals, we develop a large scale spatio-temporal binomial regression model that has both GPP and CAR components. The former is used to capture spatially varying trends by treating the trend coefficient as a non-parametric surface over the spatial domain of interest, while the latter accounts for spatio-temporal correlation. Through data augmentation steps and the use of a novel sampling strategy, we establish a modeling framework that is computationally scalable to large non-Gaussian spatio-temporal data sets. In particular, straightforward Gibbs sampling is facilitated via a data augmentation step involving latent P\'{o}lya-Gamma random variables. To avoid computationally expensive matrix calculations, we use a chromatic sampling strategy in the Gibbs sampler. Our proposed methodology easily handles missing data. The finite sample properties of our proposed approach is studied via simulation before our \textit{B.~burgdorferi} seroprevalence analysis.

The remainder of this paper is organized as follows: Section \ref{Sec:Model} describes the model and our GPP and CAR structures. Section \ref{Sec:PosteriorSampling} discusses model fitting procedures, emphasizing computational tractability with large scale spatio-temporal data. Section \ref{Sec:Simulations} presents a simulation study supporting our proposed approach, and Section \ref{Sec:Results} analyzes the canine serology data described above. We offer concluding remarks in Section \ref{Sec:Discussion}.

\section{Modeling Methods}
\label{Sec:Model}
Let $Y_{st}$ denote the number of cases (e.g., positive $\textit{B.~burgdorferi}$ tests) observed in $n_{st}$ tests in region $s$ at time $t$, for $s=1, \ldots ,S$ and $t=1, \ldots,T$.  We let  $\bs{Y}_s=(Y_{s1}, \ldots ,Y_{sT})'$, $\bs{Y}=(\bs{Y}_1', \ldots ,\bs{Y}_S')' \in \mathbb{R}^{ST}$, $\bs{n}_s = (n_{s1}, \ldots, n_{sT})'$, and $\bs{n} = (\bs{n}_1', \ldots, \bs{n}_S')' \in \mathbb{N}^{ST}$. In addition to the disease surveillance data, covariates $Z_{stq}$ and $X_{stp}$, for $q=1, \ldots , Q$ and $p=1, \ldots , P$, are assumed to be available. The $Z_{stq}$ are covariates whose associated effects are constant over the study area, while $X_{stq}$ are covariates whose associated effects vary by region.

To relate the observed test data to the available covariates, a Bayesian generalized linear mixed model \citep{McCullNelder89, DiggleEtAl98, BanerjeeEtAl04} is adopted. The general model for our data is a binomial regression: $Y_{st} | n_{st}, p_{st} \sim \text{Binomial}(n_{st},p_{st})$ with
\begin{equation}
\label{eq:LinearPredictor}
    \nu_{st} := g^{-1}(p_{st}) =  \bs{Z}_{st}'\bs{\delta} + \bs{X}_{st}'\bs\beta(\boldsymbol{\ell}_s) + \xi_{st}; ~~s= 1, \ldots, S; ~t= 1, \ldots, T,
\end{equation}
where $g:\mathbb{R} \rightarrow (0,1)$ is a known link function (e.g., logistic) relating the linear predictor $\nu_{st}$ to the prevalence $p_{st}$, $\bs{Z}_{st} = (1, Z_{st1}, \ldots, Z_{stQ})' \in \mathbb{R}^{Q+1}$, $\bs{X}_{st} = (X_{st1}, \ldots, X_{stP})' \in \mathbb{R}^P$, $\bs\delta= (\delta_0, \ldots, \delta_Q)'$ are global regression coefficients, $\bs\beta(\cdot) = (\beta_1(\cdot), \ldots, \beta_P(\cdot))'$ are spatially varying regression coefficients, $\boldsymbol{\ell}_s=(\ell_{s1},\ell_{s2})'$ is a vector of spatial coordinates (e.g., latitude and longitude) that identify the centroid of region $s$, and $ \xi_{st} $ is a spatio-temporal random effect. Following \citet{Gelfand03}, the spatially varying regression coefficients are regarded as unknown surfaces over the study region. To model these unknown surfaces while maintaining computational tractability, we use GPPs.

A Gaussian process is a stochastic process whose finite dimensional distributions are multivariate normal.  A GP $\beta_p(\cdot) \mid \bs\theta_p \sim \mathcal{GP}(\mu_p(\cdot), C(\cdot, \cdot; \bs\theta_p))$ is uniquely determined by its mean and covariance, $\mu_p(\boldsymbol{\ell}_{s}):=E[ \beta_p(\boldsymbol{\ell}_{s})]$ and $C(\bs{\ell}_s, \bs{\ell}_{s'}; \bs\theta_p) := \textrm{Cov}(\beta_p(\boldsymbol{\ell}_{s}), \beta_p(\boldsymbol{\ell}_{s'})) = \sigma_p^{2} \rho_p(\boldsymbol{\ell}_{s},\boldsymbol{\ell}_{s'}; \boldsymbol{\theta}_p)$, where $\rho_p(\cdot, \cdot ;\boldsymbol{\theta}_p)$ is a correlation function depending on the parameter vector $\boldsymbol{\theta}_p$. For smoothing and interpolation, it is often sufficient to take a constant mean \citep{BayarriEtAl07}. In our case, we \emph{a priori} posit that $\mu_p(\cdot) \equiv 0$ for all $p$. Thus,  $\boldsymbol{\beta}_{p}=(\beta_p(\boldsymbol{\ell}_1), \ldots , \beta_p(\boldsymbol{\ell}_S))',~S \in \mathbb{N},$ follows a multivariate normal distribution with mean $\bs{0}$ and covariance matrix $\bs{C}_p=\sigma_p^{2} \bs{R}_p$, where $(\bs{R}_p)_{ss'} = \rho_p(\boldsymbol{\ell}_s, \boldsymbol{\ell}_{s'};\boldsymbol{\theta}_p)$. In general, the covariance matrix inversions and factorizations associated with estimating posterior GPs are ${\cal O}(S^3)$ in computational complexity and in an MCMC context, these operations are repeated thousands of times.
Thus, as $S$ grows large, GP's quickly become computationally prohibitive.

To reduce the dimension of the problem, the Gaussian predictive process \cite[GPP;][]{Banerjee08} considers a ``parent" process based on a strategically chosen set of knots, and then interpolates the process to the points of interest via kriging. Let $\{\boldsymbol{\ell}_1^*, \ldots , \boldsymbol{\ell}_{S^*_p}^* \}$ denote the knot set with $S^*_p \ll S$. Define $\boldsymbol{\beta}_p^*=(\beta_p(\boldsymbol{\ell}_1^*), \ldots, \beta_p(\boldsymbol{\ell}_{S^*_p}^*))'$ and note that $\boldsymbol{\beta}_p^* |\sigma^2_p,\boldsymbol{\theta}_p \stackrel{ind}{\sim} \textrm{N}(\bs{0},\bs{C}_p^*)$, for all $p$, where $\bs{C}_p^*=\sigma_p^{2} \bs{R}_p^*$ and $(\bs{R}_p^*)_{ss'} = \rho_p(\boldsymbol{\ell}_s^*,\boldsymbol{\ell}_{s'}^*;\boldsymbol{\theta}_p)$. The joint distribution of $\boldsymbol{\beta}_p$ and $\boldsymbol{\beta}_p^*$ is again multivariate normal:
\begin{equation}
\label{relat}
    \left( \begin{matrix} \boldsymbol{\beta}_p \\ \boldsymbol{\beta}_p^* \end{matrix}   \right) \bigg|\sigma^2_p,\boldsymbol{\theta}_p  \sim \textrm{N}\left(\bs{0}, \sigma^2_p \left[\begin{matrix} \bs{R}_p & \widetilde{\bs{R}}_p^* \\ {\widetilde{\bs{R}}_p^{*'}} & \bs{R}_p^*   \end{matrix}\right] \right),
\end{equation}
where $\widetilde{\bs{R}}_p^*$ is an $S \times S_p^*$ matrix with the $(s,s^\prime)$th element being $ \rho_p(\boldsymbol{\ell}_s , \boldsymbol{\ell}_{s'}^*; \boldsymbol{\theta}_p)$. Exploiting this relationship, the Gaussian predictive process simply replaces $\boldsymbol{\beta}_p$ with $\widetilde{\boldsymbol{\beta}}_p := E(\boldsymbol{\beta}_p \mid \boldsymbol{\beta}_p^*; \boldsymbol{\theta}_p) = \widetilde{\bs{R}}_p^* (\bs{R}_p^*)^{-1} \boldsymbol{\beta}_p^*$.  When $S_p^*$ is not large, $(\bs{R}_p^*)^{-1}$ can be quickly computed.  For more on GPPs, see \citet{Banerjee08}.

Fully specifying a GPP requires selecting knot locations. \citet{Banerjee08} discuss several methods of knot selection, including placing them on a regular grid, selecting them at random from the observation locations, and methods which place more knots in areas with more observations. \citet{Finley08} suggest choosing knot locations to minimize the conditional variance at observed data locations; \citet{Guhaniyogi11} propose an adaptive knot selection strategy where the knot locations are treated as a point process. Following \citet{Eidsvik12}, our knots are chosen via $K$-means clustering with $S_p^*$ clusters; i.e., using $K$-means clustering we partition the $S$ counties into $S^*_p$ clusters based on the $\boldsymbol{\ell}_{s}$. The knot locations are subsequently taken to be the centroids of the $S^*_p$ clusters. For further details on $K$-means clustering see \citet{Hartigan79}.

We account for the spatio-temporal dependence that exists in the data by allowing a spatial GMRF to evolve over time. One way of doing so is presented in \cite{Waller97}, who allow the variance of the CAR model to depend on time. However, we use a first-order vector autoregression with errors following a GMRF; i.e.,
\begin{align} \label{ST:model}
    \bs \xi_t = \zeta \bs \xi_{t-1} + \bs \phi_t
\end{align}
where  $\bs \xi_t = (\xi_{1t},...\xi_{St})'$, $\zeta\in (-1,1)$ is a temporal correlation parameter, and we assume $\bs \xi_0=\bs{0}$ without loss of generality. We take $\bs \phi_t$ to be independent and identically distributed as a proper intrinsically autoregressive model \citep{BesagKoop}; i.e., $\bs \phi_t \sim \text{N} \left( \bs 0, \tau^2(\bs{D} - \omega \bs{W})^{-1} \right)$, where $\tau^2>0$ and $\omega \in (0,1)$ is a so-called `propriety parameter' that ensures the precision matrix is non-singular \citep{BanerjeeEtAl04}.  The neighborhood matrix $\bs{W} \in \mathbb{R}^{S\times S}$ is such that $(\bs{W})_{ss'}$ is equal to 1 if and only if location $s$ is adjacent to location $s', ~s \neq s'$, 0 otherwise, and $\bs{D} = \text{diag}\left(\sum_{j=1}^S(\bs{W})_{sj}, ~s= 1, \ldots, S\right)$. To avoid confounding with the intercept, we impose the standard sum-to-zero constraint (i.e., $\sum_{t=1}^T\sum_{s=1}^S \xi_{st}=0$).

We complete the proposed model by specifying prior distributions on the regression coefficients and the variance and correlation parameters.  In the absence of strong prior information, the hyperparameters are chosen so that the prior distributions are vague.  A Gaussian prior is taken on the global regression coefficients and inverse Gamma (IG) priors on the variance components for conditional conjugacy. Likewise, a truncated Gaussian prior whose support is confined to $(-1,1)$ is specified for $\zeta$, again for conditional conjugacy. We take a $\text{Beta}(\alpha_{\omega}, \upsilon_{\omega})$ prior on $\omega$ and concentrate it close to one, since previous empirical work has shown that $\omega \approx 1$ is necessary to induce noticeable spatial association \citep{BanerjeeEtAl04}. These specifications lead to the following hierarchy:
\begin{equation}\label{eqn:model}
    \begin{aligned}
  Y_{st}| n_{st}, \nu_{st} & \stackrel{indep.}{\sim} \textrm{Binomial}\left(n_{st},p_{st}=g(\nu_{st})\right), ~~s= 1, \ldots, S; ~t= 1,\ldots, T;\\
    \bs{\beta}_p^* | \sigma_p^2, \bs{\theta}_p &\stackrel{indep.}{\sim} \textrm{N}(\bs{0}, \sigma_p^2\bs{R}_p^*(\bs{\theta}_p)), ~~p= 1, \ldots, P; \\
    \sigma^2_p & \stackrel{indep.}{\sim} \textrm{IG}(\alpha_{\sigma^2_p},\upsilon_{\sigma^2_p}), ~~p= 1, \ldots, P;\\
    \bs \theta_p & \stackrel{i.i.d.}{\sim} \pi(\bs \theta_p), ~~p= 1, \ldots, P;\\
    \boldsymbol{\delta} &\sim \textrm{N}(\bs{0}, \sigma_{\delta}^2\boldsymbol{I}), ~~\sigma_{\delta}^2 > 0; \\
     \bs \xi_t| \bs{\xi}_{t-1}, \tau^2, \omega, \zeta &\sim \textrm{N}\left(\zeta \bs \xi_{t-1}, \tau^2(\bs{D}- \omega \bs{W})^{-1} \right), \quad t = 1, \ldots, T;\\
     \tau^2 &\sim \textrm{IG}(\alpha_{\tau^2},\upsilon_{\tau^2}), ~~\alpha_{\tau^2}, \upsilon_{\tau^2} > 0;\\
     \omega &\sim \textrm{Beta}(\alpha_{\omega},\upsilon_{\omega}), ~~\alpha_{\omega},\upsilon_{\omega} > 0;\\
     \zeta &\sim \textrm{Truncated Normal} (0,\sigma^2_{\zeta},-1,1), ~~\sigma^2_{\zeta} > 0,
\end{aligned}
\end{equation}
where $\nu_{st}=\bs{Z}_{st}' \bs{\delta} +\bs{X}_{st}'\widetilde{\bs\beta}(\bs{\ell}_s) + \xi_{st}$, $\widetilde{\bs\beta}(\bs{\ell}_s) = (\widetilde{\beta_1}(\bs{\ell}_s), ..., \widetilde{\beta}_P(\bs{\ell}_s))'$, and each coefficient in $\widetilde{\bs\beta}(\bs{\ell}_s)$ is obtained from the $P$ predictive processes via $\widetilde{\bs \beta}_p = \widetilde{\bs{R}}_p^*(\bs{R}_p^*)^{-1} \bs{\beta}_p^*$, and $\bs{\xi}_0 = \bs{0}$. Appropriate (identical) priors for $\bs{\theta}_1, \ldots \bs\theta_P$ depend on the selected correlation function in the GPP model.

\section{Posterior Sampling}
\label{Sec:Sampling}
\label{Sec:PosteriorSampling}
\subsection{Data Augmentation}
We assume conditional independence given the covariate effects and spatio-temporal effects and observe that $\bs Y$ depends on the regression coefficients and random effects only through $\bs{\nu} = (\nu_{11}, \ldots ,\nu_{1T}, \nu_{21}, \ldots, \nu_{ST})'$. Hence, the likelihood can be expressed as
\begin{align}\label{datamodel}
f(\bs{Y}| \bs\nu) \propto \prod_{t=1}^T \prod_{s=1}^{S} g(\nu_{st} )^{Y_{st}} \{1-g(\nu_{st}) \}^{n_{st}-Y_{st}}.
\end{align}
To develop a posterior sampling algorithm, we take $g(\cdot)$ to be the logistic link function. Other link functions are possible and can be implemented following \citet{Albert93} or \cite{Gamerman97}. Metropolis-Hastings steps \citep{MetropolisEtAl53, Hastings70} can be used either componentwise or in blocks, but such samplers can be difficult to tune in high dimensions. To facilitate the derivation of a Gibbs sampler for the regression coefficients and spatio-temporal random effects, we use a data augmentation scheme that leads to sampling these parameters from Gaussian full conditional distributions.

Our data augmentation approach follows that of \cite{Polson13}. This scheme relies on the fact that $\exp(\nu)^a\{1 + \exp(\nu)\}^{-b} = 2^{-b} \exp(\kappa \nu) \int_0^\infty \exp(- \psi \nu^2/2)$ $p(\psi|b,0) d \psi$, where $a \in  \mathbb{R}$, $b \in \mathbb{R}^+$, $\kappa = a - b/2$, and $p( \cdot \mid b,0)$ is the probability density function of a P\'{o}lya-Gamma random variable with parameters $b$ and $0$. Thus, under the logistic link, \eqref{datamodel} can be written as
\begin{align*}
    f(\bs{Y}| \bs\nu) \propto & \prod_{t=1}^T \prod_{s=1}^{S} \exp(\kappa_{st}\nu_{st}) \int_0^\infty \exp(- \psi_{st} \nu_{st}^2/2) p(\psi_{st}|n_{st},0) d \psi_{st}\\
    \propto & \prod_{t=1}^T \prod_{s=1}^{S} \int_0^\infty f(Y_{st}, \psi_{st} \mid \nu_{st})d \psi_{st},
\end{align*}
where $\kappa_{st}=Y_{st}-n_{st}/2$. By introducing the $\psi_{st}$ as latent random variables to be sampled via MCMC, we obtain
\[
f(\bs{\Y}, \bs{\psi} \mid \bs\nu) \propto \exp (-\bs{\nu'}\bs{D}_{\bs{\psi}} \bs{\nu}/2 +\bs{\kappa}' \bs{\nu} )
\prod_{t=1}^T \prod_{s=1}^{S} p(\psi_{st}|n_{st},0),
\]
where $\bs{\psi} = (\psi_{11}, \ldots, \psi_{1T}, \psi_{21}, \ldots , \psi_{ST})'$, $\bs{D}_{\bs{\psi}} = \text{diag}(\bs{\psi})$, and $\bs{\kappa} = (\kappa_{11}, \ldots , \kappa_{1T}, \kappa_{21}, \ldots, \kappa_{ST})'$. We see, then, that data augmentation yields a Gaussian density in $\bs{\nu}$ up to a normalizing constant. Consequently, the full conditional distributions for most of the parameters are of a known form and are easy to sample from; i.e., the full conditional distribution of $ \psi_{st}$ is P\'{o}lya-Gamma, $\bs \beta_p^*$ is multivariate normal, $\bs{\delta}$ is multivariate normal, $\sigma^2_p$ is inverse gamma, $\tau^2$ is inverse gamma, and $\zeta$ is truncated normal. The Supplementary Material provides the complete set of full conditional distributions.

Given the data augmentation, a posterior sampling algorithm involving Gibbs steps for the aforementioned parameters can be constructed in the usual manner. Metropolis-Hastings steps are used to sample $\bs{\theta}_p$ and $\omega$. Under the considered data augmentation scheme, the full conditional distribution of $\bs{\xi}_t$ is multivariate normal. However, sampling this parameter is computationally expensive due to its high dimension. To facilitate more efficient repeated updates of $\bs{\xi}_t$, we employ chromatic sampling, which is described next.

\subsection{Chromatic Sampling}

The full conditional distributions of $\bs{\xi}_t, ~t=1, \ldots ,T$, are each multivariate normal. Block sampling from these full conditionals is reasonable when the number of spatial units is relatively small  \citep{Furrer10}, but becomes unwieldy as $S$ increases due to the associated Cholesky factorizations and memory requirements. As an alternative, we propose to use so-called {\em chromatic sampling} \citep{Gonzalez11, Brown17}. The chromatic sampler exploits the Markov structure of the CAR model to parallelize single-site updates, thereby avoiding time consuming matrix calculations such as Cholesky factorizations. Under chromatic sampling, the computing time scales approximately linearly in $S$ and $T$.

Let $\{\mathcal{A}_1, \ldots, \mathcal{A}_K\}$ be a partition in which $\mathcal{A}_k$ is an index set identifying a collection of spatial regions that are not adjacent to one another; i.e., for all $s,s'\in \mathcal{A}_k$, $\bs{W}_{ss'}=0$. A greedy algorithm for finding such a partition on an irregular lattice is given by \citet{Brown17}. For a vector $\bs{a} = (a_1, \ldots, a_S)'$ and an index set $\mathcal{C}$, define $\bs{a}(\mathcal{C}):= (a_s : s \in \mathcal{C})'$. The Markov property of the CAR model implies that the elements of $\bs{\xi}_{t}(\mathcal{A}_k)$, given $\bs{\xi}_{t}(\mathcal{A}_k^c)$, are conditionally independent. Therefore, by conditioning on $\bs{\xi}_{t}(\mathcal{A}_k^c)$, the elements of $\bs{\xi}_{t}(\mathcal{A}_k)$ can be sampled from their univariate full conditional distributions in parallel (or through `vectorized' calculations). This approach can handle an extremely large number of spatial regions (e.g., $S>100,000$) when they are sparsely connected. For further details, see \cite{Brown17}, who compare block sampling to chromatic sampling for GMRFs.

\subsection{A Note on Missing Data}

In our application, data are not reported at all location-time pairs. To capture this effect, let $\mathcal{R}$ be the set of ordered pairs $(s,t)$ for which data are available. The augmented likelihood is then
$$f(\Y(\mathcal{R}), \bs \psi(\mathcal R) \mid  \bs\nu(\mathcal{R}))  \propto \exp ( - \bs \nu(\mathcal{R})' \bs{D}_{\bs \psi(\mathcal{R})} \bs \nu(\mathcal{R})/2 +\bs \kappa(\mathcal{R})' \bs \nu(\mathcal{R}) ) \prod_{(s,t) \in \mathcal R} p(\psi_{st}|n_{st},0),$$
where $\bs\nu(\mathcal{R}) = \bs Z(\mathcal{R})\bs\delta + \bs{X}(\mathcal{R})\widetilde{\bs{b}} + \bs{I}(\mathcal{R})\bs \xi$ and we use the convention that $\bs A(\mathcal{R})$ is the matrix formed by retaining the rows of $\bs A$ whose indices are in $\mathcal{R}$. Here $\bs Z =(\bs Z_1' \cdots \bs Z_S')' \in \mathbb{R}^{ST \times (Q+1)}$ with $\bs Z_s = (\bs Z_{s1} \cdots \bs Z_{sT})'$. Similarly, $\bs X = \bigoplus_{s=1}^S \bs X_s \in \mathbb{R}^{ST \times SP}$ with $\bs{X}_s = (\bs X_{s1}, \ldots, \bs X_{sT})'$, $\bs{I}$ is the identity matrix, and $\widetilde{\bs b} = (\widetilde{\bs\beta}'(\bs{\ell}_1), \ldots, \widetilde{\bs\beta}'(\bs{\ell}_S))' \in \mathbb{R}^{SP}$. Since $\bs \xi \in \mathbb{R}^{ST}$ is the vector of spatial random effects over {\em all} locations within the study region for all time points, we obtain a well-defined full conditional distribution for $\bs{\xi}$, provided that the prior on $\bs{\xi}$ is proper. This representation of the joint density allows the model to be extended to the entire study region by imputing the missing effects via posterior realizations.

\section{A Simulation Study}
\label{Sec:Simulations}

In this section, we study how well the proposed method estimates model coefficients and how GPP knot selection influences results via simulation. Data are generated on a regularly spaced $13 \times 13$ grid over 60 time points and drawing $Y_{st}|n_{st},p_{st} \stackrel{indep.}{\sim} \textrm{Binomial}(n_{st},p_{st})$, where
\[
g^{-1}(p_{st}) = \delta_0 + \widetilde{\beta_1}(\bs{\ell}_s)t/60 + \xi_{st}, ~~s= 1, \ldots, 13^2; ~t= 1, \ldots, 60,
\]
and $g(\cdot)$ is the logistic link. The test counts $n_{st}$ are randomly sampled from a discrete uniform distribution ranging from 100 to 200. The random effects $\xi_{st}$ are generated from the CAR model defined in Section \ref{Sec:Model}, with $\zeta=0.9$, $\tau^2=0.005$, $\omega=0.9$, and the neighborhood matrix $\bs{W}$ set so that two areas are neighbors if and only if they share a common edge or corner. The true intercept is $\delta_0=-1$ and the surface $\widetilde{\bs{\beta}}_1(\cdot)$ at each study location is generated from a GPP model. In particular, a realization of the parent process is first simulated on a $5 \times 5$ grid of equally spaced knots. The parent process has $\mu_1(\bs{\ell}_s^*) \equiv 1$ and $\rho(\bs{\ell}_s^*, \bs{\ell}_{s'}^*; \bs{\theta}_1) = \bs{\theta}_1^{d_{ss'}^2}$, where $d_{ss'}$ is the Euclidean distance between $\bs{\ell}_s^*$ and $\bs{\ell}_{s'}^*$, $\bs{\theta}_1=0.6$ and $\sigma^2_1 = 1.5$. The resulting true surface $\widetilde{\bs{\beta}}_1(\cdot)$ is depicted in Figure 2. Using this surface, 500 independent data sets are generated from the assumed data generating model.

We fit our model to each of the 500 data sets using three separate knot set configurations. The first configuration uses the same knots as those used to generate the true surface, representing an ideal situation. The other two configurations take $4 \times 4$ and $7 \times 7$ grids of equally spaced knots. For the model \eqref{eqn:model} priors, we take $\alpha_{\sigma^2_1}= \upsilon_{\sigma^2_1} = \alpha_{\tau^2} = \upsilon_{\tau^2}=2$, $\sigma_{\delta}^2 = 1000$, $\alpha_{\omega}=900$, $\upsilon_{\omega}=100$, and $\sigma^2_\zeta=10$. In the GPP, the correlation function is taken to be $\rho(\bs{\ell}_s, \bs{\ell}_{s'};\bs{\theta}_1) = \bs{\theta}_1^{d_{s,s'}^2},$ the same as the true GPP, and we specify a $\text{Uniform}(0,1)$ prior on $\bs{\theta}_1$. For each data set, we retain 5,000 MCMC iterates after a burn-in of 5,000 samples. Convergence of the chains were assessed via trace plots.

Figure 3 displays a summary of the simulation results for the temporal trend parameter $\widetilde{\bs{\beta}}_1(\cdot)$. This summary includes a spatial depiction of the arithmetic average of the 500 point estimates, as well as empirical bias and mean squared error, where for each simulated data set a point estimate of $\widetilde{\bs{\beta}}_1(\cdot)$ was obtained as the mean of the 5,000 retained MCMC iterates. The results suggest that our model estimates the spatially varying regression coefficient well; i.e., the mean estimates show little bias. The variability of the estimators tends to increase near the region's edges. This boundary effect is expected and is common to non-parametric smoothers. Further, little difference between the estimates obtained under the three different knot configurations is seen, demonstrating that the methods can recover the true coefficient surface across the entire study region (assuming the model is correct up to choice of knots).

\section{Lyme Analysis}
\label{Sec:Results}

\subsection{Background}
\label{Sec:Motivation}

Our data consist of 16,571,562 test results from domestic dogs living throughout the conterminous United States from January 2012 - December 2016.  The data were provided by IDEXX Laboratories, Inc. to the Companion Animal Parasite Council (CAPC), who made them available online at \url{https://www.capcvet.org}. The data are aggregated by month and county, resulting in 69,876 county-month pairs reporting data.

In general, the spatial distribution of a vector-borne disease is strongly influenced by the environment and the vector's hosts, leading to correlated data \citep{Legendre1993}. Indeed, a strong spatial correlation is seen in these data, as indicated by Figure 1 and a Moran's $I$ statistic of 0.378 ($p$-value $\approx 0$). Such data are also positively temporally correlated. Figure 4 displays raw county-level seroprevalence estimates over all months in the respective years of 2012 and 2016. These figures suggest where a significant increase in seroprevalence is expected and include western Pennsylvania, Virginia, West Virginia, Minnesota, and Iowa.

\subsection{Model Building and Seasonality}
Given the seasonality of \textit{Ixodes} spp.~activity, seasonality could manifest itself in \textit{B.~burgdorferi} seroprevalence. To investigate this, the model
\begin{equation}
\label{eq:SpatialSeasonalModel}
    \nu_{st} = \delta_0  + \widetilde{\beta}_1(\bs{\ell}_s)I_1(t) + \widetilde{\beta}_2(\bs{\ell}_s)I_2(t) + \widetilde{\beta}_3(\bs{\ell}_s)I_3(t) + \widetilde{\beta}_4(\bs{\ell}_s)t + \xi_{st}
\end{equation}
is posited, where $t$ denotes time (rescaled to the unit interval) and $I_p(t)$ is a seasonal indicator, for $p=1,2,3$. Seasons are defined as follows: winter (December-February), spring (March-May), summer (June-August), and fall (September-November), where winter regarded as the baseline. This model allows for spatially varying seasonal effects and spatially varying trend effects. While covariates such as county level temperatures and precipitations are available, these are not used in the regression since our goal is to quantify any trends, not determine specific drives of any trends.

The model in \eqref{eq:SpatialSeasonalModel} was fit with the prior specifications and correlation function described in Section 4. Two specifications for the GPP model are considered, using 50 and 100 knots, respectively. In both cases, knot placement for all GPP models is done by K-means clustering as described in Section 2.1. For sampling, 30,000 MCMC iterates are generated, with the last 10,000 retained for inference. Convergence of the MCMC chains was assessed using trace plots. We stress the computational scalability of this approach.  This model consists of four {\em a priori} independent coefficient surfaces, each with 3109 spatial locations, and 186,540 spatio-temporal random effects.

Two primary findings arise. First, there are no appreciable differences between the estimates using 50 and 100 knots. As both specifications are computationally feasible, all subsequent analyses use 100 knots. Second, there is evidence of seasonality in the location parameters, but these appear constant across space. Thus, the simpler model
\begin{equation}
\label{eq:ConstantSeasonalModel}
    \nu_{st} = \delta_0 + \delta_1I_1(t) + \delta_2I_2(t) + \delta_3 I_3(t) + \widetilde{\beta}_1(\bs{\ell}_s)t + \xi_{st}
\end{equation}
was fit. Credible intervals at level 95\% indicate that the model can be futher reduced to
\begin{equation}
    \label{eq:WinterVersusNonwinter}
    \nu_{st} = \delta_0 + \delta_1 I_{1}^*(t)  + \widetilde{\beta}_1(\bs{\ell}_s)t + \xi_{st},
\end{equation}
where $I_{1}^*(t)$ is a seasonal indicator that equals one if $t$ is between March and November, and zero otherwise. Approximate 95\% credible intervals for $\delta_0$ and $\delta_1$ are $[-3.95, -3.82]$ and $[-0.20, -0.10]$, respectively.

For further insight, the model in \eqref{eq:WinterVersusNonwinter} is compared against the nonseasonal model
\begin{equation}
    \label{eq:NonSeasonal}
     \nu_{st} = \delta_0 + \widetilde{\beta}_1(\bs{\ell}_s)t + \xi_{st}.
\end{equation}
For this model, an approximate 95\% credible interval for $\delta_0$ is $[-4.08, -4.03]$. Figure 5 displays the estimates of $\widetilde{ \beta}_1(\cdot)$ from both models. Very similar large-scale patterns in the estimated trends are seen; hence, while seasonality exists in the location parameters, its effect on trends seems negligible.


The spatial trend $\widetilde{\beta}_1(\cdot)$ is a large-scale regional trend with low spatial frequency, and is not intended to explain local (county-level) trends. While regional trends are useful for estimating behavior in areas reporting little data, it may be desirable to separate local heterogeneity in the trends and provide a county level assessment. Our proposed modeling framework can accomplish this.  Specifically, let $\upsilon_s^{(g)}$, for posterior sample realization $g=1,...,G$, be the slope estimate obtained at county $s$ by fitting a simple linear regression to $\{(t,\nu^{(g)}_{st}):~t=1,...,T \}$.   Then $\upsilon_s^{(g)}$ can be regarded as a Monte Carlo realization of the county level trend. Using the $\upsilon_s^{(g)}$ as a random sample, point estimation and inference for county level trends proceeds in the usual way.

\subsection{Results}

Figure 5 displays the estimated posterior mean of the regional trend $\widetilde{\bs \beta}_1$.  The regional rate of change in \textit{B.~burgdorferi} seroprevalence between January 2012 and December 2016 is positive in all states that are currently recognized as having high human Lyme disease incidence \citep{CDClymestats}, including portions of the Northeast and the Upper Midwest. The regional rate of increase varies spatially, with high incidence regions generally exhibiting the greatest changes. These regions include Maine, south to West Virginia and Virginia, and northern parts of Minnesota and Wisconsin.

Figure 6 displays estimated posterior means of the county-level trend $\upsilon_s$.  Significantly increasing local trends are seen in much of the Northeast, extending southwards through West Virginia and Virginia and into North Carolina and Tennessee.  This conclusion is not surprising as this region entails localities where Lyme disease has been reported in increasing incidence.  Also seen are increasing local trends in parts of northwestern Minnesota, northern Wisconsin, and southeastern Iowa. In the Great Lakes region, increasing trends are observed in eastern Ohio, Indiana, and western Michigan.  Of note, in much of eastern New England which is the region where Lyme first emerged in people, the prevalence appears to be remaining stable, albiet high.  Figure 7 graphically depicts counties where local trends are significantly positive, using approximate 95\% equal-tailed credible intervals to assess significance.  This graphic further supports the above statements.

\section{Discussion}\label{Sec:Discussion}

This paper develops a computationally feasible binomial regression model for a large spatio-temporal data set that can identify localized trends in canine seroprevalence. Our novel approach combines several recent advances in large-scale spatial modeling and MCMC sampling. The end product is a flexible, scalable methodology for modern spatio-temporal data.

Our approach was used to identify regions of the U.S. experiencing increasing canine risk for \textit{B.~burgdorferi} infection.  Since human and canine risk are similar, these regions are likely experiencing increasing human exposure as well. And while human Lyme disease data may be private and in many regions scarce due to lack of testing, our canine seroprevalence data consist of over 16 million spatio-temporally referenced test results. The size of the domain (3109 spatial locations and 60 time points) creates computational challenges. While monthly and county-level aggregation reduces the size of the response vector from 16,581,562 test results to 69,876 county-month pairs, a binomial response in an MCMC context typically requires sampling via Metropolis-Hastings steps, which can be difficult to tune in extremely high dimensions (over 180,000, in our case). Under the logistic link, a recently proposed Poly\'{a}-Gamma data augmentation was used to facilitate direct Gibbs sampling on full conditional distributions. Gaussian predictive processes were used to model smoothly varying, high-dimensional coefficients through a low-dimensional representation. Local spatio-temporal heterogeneity was modeled by random effects following a time-varying Gaussian CAR distribution. Chromatic sampling was used on GMRFs to construct an efficient MCMC algorithm.

The motivation for this study is the rise in reported Lyme disease cases in the United States \citep{Adams2017} and, in particular, rising incidence in states not traditionally considered to be endemic, such as West Virginia. Our results suggest that 1) canine seroprevalence is rising in conjunction with reports of human cases \citep{Kugeler2015, Hendricks2017,CDClymestats}, 2) rates are increasing most in areas where the pathogen has recently encroached, and 3) seroprevalence in dogs is rising outside of the states considered to be high incidence for humans \citep{CDClymestats} (suggesting that risk may be increasing for humans in those areas). Several studies have recognized increasing risk in low incidence areas, including human disease incidence, tick density, and presence of the pathogen. These areas include Illinois \citep{Herrman2014}, Iowa \citep{Lingren2005}, North Dakota \citep{Russart2014}, Ohio \citep{Wang2014}, and Michigan \citep{Lantos2017}. We also observe significant increases in canine seroprevalence in several states that have not yet reported significant human incidence. Given the proximity of these areas to recognized high-incidence states, it is reasonable to propose that canine seroprevalence is more sensitive to changes in risk of exposure and thus may be used as an early warning system for changes in human risk.

Examining local trends, as opposed to regional effects, shows that some adjacent counties are exhibiting trends in opposite directions. To fully understand this heterogeneity, further ecological analyses are needed. Possible factors to consider include the presence of urban centers, degree of forestation or other habitat factors, tick populations, reservoir presence and densities, vaccination, and preventative medication use in dogs. The latter are likely driven by socioeconomic factors whereas other factors are related to climate or changing habitats. Areas with significantly positive trends include the Appalachian mountains from upstate New York to North Carolina, the Upper Midwest, and Iowa. The West Virginia, western Pennsylvania, and eastern Ohio regions can be viewed as a leading edge of rising seroprevalence in Lyme's westward expansion. This is supported by evidence in increased reports of ticks in these regions \citep{Eisen2016tick}.

Our approach makes several simplifying assumptions. We treat the link function in the GLM as known, and it might be poorly specified. As this can induce bias in the estimates of the covariate effects \citep{Neuhaus99}, relaxing this assumption could be fruitful. We also assumed that the spatially varying coefficients follow independent and identically distributed Gaussian processes. A more flexible approach would allow these coefficients to be correlated through a multivariate GP \citep{VerHofBarry98}, but these are more difficult to use and challenges remain in their development \citep[e.g.,][]{FrickerEtAl13}.  The observed seroprevalences suggest that smoothness of the random effects may change by region, suggesting that a heteroskedastic GP might be more appropriate \citep{BinEtAl16}. Further, GMRFs are known to oversmooth salient features \citep{SmithFahr07} and do not directly correspond to any valid covariance function in a GP. However, approximating GPs with GMRFs via stochastic PDEs to maintain computational feasibility \citep{Lindgren11} could be promising for our application. In addition to statistical challenges, future applications of our model include human Lyme disease data and heartworm disease, ehrlichiosis, and anaplasmosis in canines. The ecological, entomological and environmental implications of the canine Lyme seroprevalence analysis presented in this work is the subject of ongoing research.

\section{Supplementary Material}
The supplementary material for this article includes the full conditional distributions required to develop the proposed posterior sampling procedure.

\section*{Acknowledgements:} The authors thank IDEXX Laboratories, Inc.~for their data contribution. This material is based upon work partially supported by the National Science Foundation Grants DMS-1127914, DMS-1407480, CMMI-1563435, and EEC-1744497, and National Institutes of Health Grant R01 AI121351. JRG is supported by The Boehringer Ingelheim Vetmedica-CAPC Infectious Disease Postdoctoral Fellowship.

\bibliography{references}

\newpage

\begin{center}
    \huge{Supplementary Material for ``A Large Scale Spatio-temporal Binomial Regression Model for Estimating Seroprevalence Trends"}\\[18pt]

    \large{Stella Watson Self, Chrisopher McMahan, D. Andrew Brown, Jenna Gettings, and Michael Yabsley}
\end{center}

    \setcounter{equation}{0} 
\section*{Appendix A: Full Conditional Distributions}\label{Sec:Appendix1}

This section provides the full conditional distributions of all unknown parameters from the model described in Section 2 of the paper. To provide for the most general setting possible, these derivations account for the potential of missing data throughout time and space, through adopting the notational conventions depicted in Section 3.3. Using this notation we now depict the full conditional distributions of all of the model parameters.

For ease of exposition and clarity in the following derivations, the dependencies in the notational depiction of the full conditional distributions are suppressed, with the parameters of these distributions implicitly identifying these quantities. The full conditional distribution of the latent $\psi_{st}$'s is P\'{o}lya-Gamma$(n_{st}, \nu_{st})$. For further details on the structure of the P\'{o}lya-Gamma distribution as well as sampling strategies see \cite{Polson13}. Exploiting the normal form in $\bs{\nu}$, it is easy to establish that the full conditional distribution of $\bs \b_p^*$ is $\text{MVN} ( \bs \mu_{\bs \b_p^*}, \bs \Sigma_{\bs \b_p^*} )$, where
\begin{align*}
 \bs \mu_{\bs \b_p^*} &= \bs\Sigma_p \left \{ \bs T_p( \mathcal R)'\bs X_p(\mathcal R)' \bs \kappa - \bs T_p( \mathcal R)'\bs X_p(\mathcal R)' \bs D_{\bs \psi}\bs \nu_{\bs \b_p^*} \right\}, \\
 \bs \Sigma_{\bs \b_p^*} &= \left \{ \bs T_p(\mathcal R)' \bs X_p(\mathcal R)'\bs D_{\bs \psi} \bs X_p (\mathcal R)\bs T_p(\mathcal R)  + \bs C_p^{*-1} \right\}^{-1}.
\end{align*}
Note, in the expression above we have that $\bs{T}_p$ is created by stacking the matrix $\widetilde{\bs{R}}_p^*(\bs{R}_p^*)^{-1}$ on itself $T$ times, $\odot$ denotes the Hadamard product, and
$$\bs \nu_{\bs \b_p^*} =  \bs Z({\mathcal R}) \bs \delta +  \sum_{p'\neq p} \bs X_{p'}( \mathcal R) \odot \bs{T}_{p'}( \mathcal R)\bs\beta_{p'}^* + \bs I(\mathcal R) \bs \xi.$$
Similarly, the full conditional distribution of $\bs \delta$ is MVN$(\bs \mu_{\bs \delta}, \bs \Sigma_{\bs \delta})$,
where
\begin{align*}
\bs \mu_{\bs \delta} &= \bs \Sigma_{\bs \delta}\left\{ \bs Z(\mathcal R)' \bs \kappa - \bs Z(\mathcal R)' \bs D_{\bs \psi} \bs \nu_{\bs \delta}  \right\}, \\
\bs \Sigma_{\bs \delta} &=  \left\{ \bs Z(\mathcal R)'\bs D_{\bs \psi} \bs Z(\mathcal R)+\sigma_{\bs \delta}^{-2}\bs I \right\}^{-1},
\end{align*}
and here
$$\bs \nu_{\bs \delta} =  \sum_{p=1}^P \bs X_{p}( \mathcal R) \odot \bs{T}_{p}( \mathcal R)\bs\beta_{p}^* + \bs I(\mathcal R) \bs \xi.$$

\noindent The full conditional distribution of $\sigma_p^{2}$ is Inverse Gamma$ \left ( S_p^*/2 + \a_{\sigma^2_p},  \bs \beta_p^* \bs R_p^{*-1}\bs \beta_p^*/2  + \upsilon_{\sigma^2_p} \right)$ and the full conditional distribution of $\tau^2$ is Inverse Gamma$(\alpha^*_{\tau^2}, \beta^*_{\tau^2})$, where
\begin{align*}
    \alpha^*_{\tau^2} &=  TS/2 + \alpha_{\tau^2},\\
    \beta^*_{\tau^2} &=   \frac{1}{2} \left( \bs \xi_1^T\{ \bs D - \omega \bs W\} \bs \xi_1 + \sum_{t=2}^T \{\bs \xi_t -  \zeta \bs \xi_{t-1}\}^T\{\bs D - \omega \bs W\} \{\bs \xi_t - \zeta \bs \xi_{t-1}\} \right)+ \upsilon_{\tau^2}.
\end{align*}

\noindent The full conditional distribution of $\zeta$ is Truncated Normal$(\mu^*_{\zeta}, \sigma^*_{\zeta}, -1,1)$ where
\begin{align*}
\mu^*_{\zeta}       &= \sigma^{2*}_{\zeta} \sum_{t = 2}^T \bs \xi_{t-1}' \bs A\bs \xi_t ,  \\
\sigma^{2*}_{\zeta} &= \left ( \sum_{t = 2}^T \bs \xi_{t-1}' \bs A \bs \xi_{t-1} + \sigma_\zeta^{-2} \right)^{-1} ,
\end{align*}
and $\bs A = \tau^{-2}(\bs D - \omega \bs W)$. Attention is now turned to the spatio-temporal random effects. The full conditional distribution of $\bs \xi_t$ is MVN$(\bs \mu_{\bs \xi_t}, \bs \Sigma_{\bs \xi_t})$, where
\begin{align*}
\bs \mu_{\bs \xi_t} & = \bs \Sigma_{\bs \xi_t} \left \{  \bs I(\mathcal R_t)'\bs \kappa_t -\bs I(\mathcal R_t)'\bs D_{\bs \psi_t} \bs \nu_{\bs \xi_t} + \zeta \bs A ( \bs \xi_{t-1} + \bs \xi_{t+1}) \right \},  \\
\bs \Sigma_{\bs \xi_t} &= \left \{ \bs I(\mathcal R_t)' \bs D_{\bs \psi_t} \bs I(\mathcal R_t) + (1 + \zeta^2) \bs A \right \} ^{-1} \text{ for }t = 1, ..., T-1,\\
\bs \Sigma_{\bs \xi_t} &= \left \{ \bs I(\mathcal R_t)' \bs D_{\bs \psi_t} \bs I(\mathcal R_t) + \bs A \right \} ^{-1}, \text{for t = T}.
\end{align*}
In the above parameter configurations, we have that $\mathcal{R}_t$ is the collection of ordered pairs $(s,t)$ which denote the locations at which data was collected during the $t$th time period, $\bs I$ is an identity matrix, $\bs D_{\bs \psi_t}=\textrm{diag}\{\bs \psi(\mathcal{R}_t) \}$, $\bs \kappa_t = \Y(\mathcal{R}_t) - \bs{n}(\mathcal{R}_t)/2$, $\bs \xi_0 = \bs \xi_{T+1}=\mathbf 0$, and
$$\bs \nu_{\bs \xi_t} =  \bs Z({\mathcal{R}_t}) \bs \delta + \sum_{p=1}^P \bs X_{p}( \mathcal{R}_t) \odot \bs{T}_{p}( \mathcal{R}_t)\bs\beta_{p}^* .$$
The aforementioned full conditional distributions of $\bs \xi_t$ can be used if the number of spatial units is relatively small. When faced with a large number of areal units it is suggested that the chromatic sampler discussed in \citet{Brown17} be utilized. To accomplish this, one need only have a ``coloring" of the areal units and know the full conditional distributions of the $\xi_{st}$. A coloring of the areal units can be obtained using the algorithm developed in \citet{Brown17}, and we note that the full conditional distributions of the $\xi_{st}$
is Normal$(\mu_{\xi_{st}}, \sigma^2_{\xi_{st}})$, where

\begin{align*} \small
 \mu_{st} & = \frac{\sigma_{st}^2}{\tau^2} \left \{ \omega(1 + \zeta^2) \bs{w}_{s} \bs{\xi}_{t} -g_{st}\tau^2(\psi_{st}\nu^{*}_{st} -\kappa_{st}) + \zeta d_s ( \xi_{s,t-1} + \xi_{s,t+1}) -\zeta \omega \bs w_{s} (\bs{\xi}_{t-1} + \bs{\xi}_{t+1})  \right\}\\
 \sigma_{st}^2&  = \left \{g_{st}\psi_{st}+ \frac{d_s(1 + \zeta^2)}{\tau^2} \right\}^{-1},
\end{align*}
for $t = 1,...,T-1$, and
\begin{align*} \small
\mu_{st} & = \frac{\sigma_{st}^2}{\tau^2} \left \{ \omega \bs{w}_{s} \bs{\xi}_{t} -g_{st}\tau^2(\psi_{st}\nu^{*}_{st} -\kappa_{st}) + \zeta d_s  \xi_{s,t-1} -\zeta \omega \bs w_{s} \bs{\xi}_{t-1}   \right\}\\
\sigma_{st}^2&  = \left \{ g_{st}\psi_{st}+ \frac{d_s}{\tau^2} \right\}^{-1}.
\end{align*}
\noindent for $t = T$. Here $\bs w_{s}$ denotes the $s$th row of $\bs W$, $d_s$ denotes the $s$th diagonal element of $\bs D$ $\nu_{st}^* = \bs Z_{st} \bs \delta + \sum_{p=1}^P X_{stp} \tilde \beta_p(\bs l_s)$, and  $g_{st} = 1$ if $(s,t) \in \mathcal R$ and 0 otherwise.

As is common for Gaussian process and Gaussian predictive process models, the full conditional distribution of $\bs \theta_p $ is not a recognizable distribution but is proportional to:
\begin{align*}
& \exp \left( - \frac{1}{2} \left \{ \widetilde{\bs \beta}_p'\odot \bs X_p(\mathcal R)'\left[ \bs D_{\bs \psi} \bs X_p(\mathcal R) \odot \widetilde{\bs \beta}_p + 2 \left(\bs D_{\bs \psi} \bs \nu_{\bs \beta_p^*} -  \bs \kappa \right) \right]+ \bs \b_p^{*T}\bs C_p^{*-1}  \bs \b_p^*  \right \} \right)
\det\{\bs C_p^*\}^{\frac{-1}{2}}
\end{align*}
Recall that $\bs \theta_p$ is embedded in $\widetilde{\bs \beta}_p = \bs T_p(\mathcal R) \bs{\beta}_p^*$, where $\bs T_p$ is a function of $\bs \theta_p$. In general, a Metropolis-Hastings step is used to sample $\bs \theta_p$ and this step might depend on the specific form of the selected correlation function.

Similarly, the full conditional distribution of $\omega$ is not a recognizable distribution, but it is proportional to:
\begin{align*}
&  \exp \left\{ - \frac{1}{2} \sum_{t=1}^T ( \bs \xi_t - \zeta \bs \xi_{t-1})'\bs A (\bs \xi - \zeta \bs \xi_{t-1})  \right\} \det\{\bs A \}^{ \frac{T}{2}} \omega^{\alpha_\omega-1}(1- \omega)^{v_\omega - 1},
\end{align*}
where $\bs A = \tau^{-2}(\bs D - \omega \bs W)$. A Metropolis-Hastings step is used to sample $\omega$.


\end{document}